\documentstyle[manuscript,aps]{revtex}
\begin{document}

\title{
  Quantum limits and symphotonic states in free-mass gravitational-wave
  antennae.
}
\author{
  V.B.Braginsky, M.L.Gorodetsky, F.Ya.Khalili \\
  {\it Dept. of Physics, Moscow State University}, \\
  {\it Moscow 199899, Russia}, \\
  {e-mail: brag@hbar.phys.msu.su}
}

\maketitle

\begin{abstract}
Quantum mechanics sets severe limits on the sensitivity and required
circulating energy in traditional free-mass gravitational-wave antennas.
One possible way to avoid these restrictions is the use of intracavity QND
measurements. We analyze a new QND observable, which possesses a number of
features that make it a promising candidate for such measurements and
propose a practical scheme for the realization of this measurement.
In combination with an advanced coordinate meter, this scheme makes it
possible to lower substantially the requirements on the circulating power.
\end{abstract}

\section{Introduction}

In \cite{NL96,optbar}, we presented an analysis of two qualitatively new
schemes for the extraction of information from free-mass gravitational-wave
antennas \cite{LIGO}. Common features of these schemes are the use of nonclassical
quantum states of the optical field inside the resonators and of QND methods
for intracavity measurements of the variations of these states.  This becomes
possible only with the realization of optical field relaxation times
$\tau^*_{o}$ much longer than the measurement time
$\tau_{meas}\simeq 10^{-2} \div 10^{-3}s$. One significant advantage of
intracavity measurements is that they require lower levels of circulating
power than traditional schemes with an antenna with a coherent pump.  In
\cite{NL96} and in the subsequent article by Levin \cite{Levin}, the optical
cubic nonlinearity $\chi^{(3)}$ of thin plates inserted in an antenna was
exploited.

The idea of our second scheme \cite{optbar}, which, in our opinion, can be implemented
relatively easily, was to place an additional partially transparent
mirror--probe mass at the intersection of the two arms of a gravitational
antenna.  This results in the formation of two coupled Fabry-Perot resonators.
Displacement of the end mirrors under the action of a gravitational wave
leads to a redistribution of the energies in the arms, which pushes the central
mass.  The absolute displacement under optimal conditions is simply equal to
the relative displacements of the end mirrors ($hL/2$, where $L$ is
the arm length and $h$ is the amplitude of the variation of the metric), and
the light in the system behaves like a rigid bar.  The displacement associated
with an independent mass that does not interact with the optical field
can be registered without consuming a large amount of power.  A rigorous
general relativistic justification of the schemes in \cite{NL96,optbar} can be
found in \cite{Sazhin}.

The merits of this intracavity measurement are the following: a) in
the resonators, the required nonclassical quantum state (close to a Fock state)
is formed automatically; b) direct measurement of a displacement
$hL/2$ consumes relatively little power; c) precision higher than
the standard quantum limit can be obtained.

In \cite{optbar}, we did not make an analysis of the minimal energy of the
optical field ${\cal E}$ in the system required to preserve the sensitivity.
Another important unanalysed problem is the connection between the achievable
resolution and a chosen procedure for displacement measurement.

It is important to note here that the provision of substantial values of
${\cal E}$ is a key problem for large-scale gravitational wave antennas,
and that this problem has not a technical but a fundamental nature.
Indeed, the proposed sensitivity levels of such antennas will be close to the
standard quantum limit for the displacement of the masses $M$ of the end
mirrors:

\begin{equation}
  \Delta x_{SQL}(M) = Lh_{SQL} \simeq
    \sqrt{\frac{\hbar}{M\omega_{gr}^2\tau_{gr}}},
\end{equation}
where $\omega_{gr}$ is the frequency of the gravitational signal and
$\tau_{gr}$ is its duration (we omit in our estimates numerical terms of
the order of unity that depend on the form of the signal).  According to the
Heisenberg uncertainty relation, the momentum should be perturbed by a
value of the order of:

\begin{equation}
\Delta p = \frac{\hbar}{2\Delta x_{SQL}} \simeq
     \sqrt{\hbar M\omega^2_{gr}\tau_{gr}}
\end{equation}
This perturbation must be provided by the uncertainty in the energy
${\cal E}$ in the interferometer, which, thus, cannot be less than

\begin{equation}
  \Delta{\cal E} = \sqrt{\frac{\omega_{gr}}{\tau_{gr}}}L\Delta p =
    L\sqrt{\hbar M\omega^3_{gr}}
\end{equation}
This value is not especially large; for example, for $L=4\times10^5 cm$,
$\omega_{gr}=10^3 s^{-1}$, and $M=10^4 g$ (the parameters of the LIGO antenna),

\begin{equation}
\Delta{\cal E}\simeq 4\times 10^{-2} erg,
\end{equation}
and in the case of nonclassical states of the optical field, in which $\Delta
{\cal E}\sim {\cal E}$, the necessary resolution can be obtained at very low
energies.  However, for coherent states in which

\begin{equation}
\Delta{\cal E}=\sqrt{\hbar \omega_o{\cal E}},
\end{equation}
where $\omega_o$ is the optical frequency, the requirements are very strict:

\begin{equation} {\cal E}\simeq \frac{ML^2\omega^3_{gr}}{\omega_o}. \end{equation}

For the same parameters as before and $\omega_o=2\times 10^{15}s^{-1}$,

\begin{equation} {\cal E}_{SQL}\simeq 10^9 erg, \end{equation}
and if $\omega_{gr}=10^4s^{-1}$, then

\begin{equation} {\cal E}_{SQL}\simeq 10^{12} erg. \end{equation}

In this paper, we analyze a new intracavity scheme that is, in some sense,
complementary to the ``optical bars'' scheme.  In this scheme, the optical
field forms in a quantum state that is close to states with squeezed phase;
this is known to allow, in principle, a dramatic decrease in the optical
quanta because $\delta\varphi\sim 1/N$. (Non-QND measurement of a similar
observable was proposed in \cite{Burnett}).

\section{A crossquadrature quantum observable and a scheme for its measurement}

The basic idea of the new scheme for an intracavity readout system is the
use of two modes excited in the Fabry-Perot resonators of the antenna's
orthogonal arms. If the modes are not linearly coupled (this is critical in
this scheme), they can be tuned as close to each other as
$(\omega_1-\omega_2)\tau_{meas}\ll1$. As a result, the frequency variation
in one (or both) resonators produced by a gravitational wave will lead to the
appearance of a phase difference with the oscillation amplitude

\begin{equation}
  \delta\varphi\simeq\frac{h\omega_o}{\omega_{gr}},
\end{equation}
which we propose to register. Since no meter has been invented thus far to
directly register the phase difference between two quantum electromagnetic
oscillators, another variable proportional to $\delta \varphi$ is required.

We propose to measure the averaged product of the two quadrature components of
two different oscillators, which, in the limit of large numbers of quanta, is
very close to a phase measurement. One possible scheme for the realization of
the proposed crossquadrature observable is depicted in Fig.1. This scheme
is based on the use of ponderomotive nonlinearity in a way similar to that
in \cite{optbar}. Mirrors $A'$ and $B'$ direct the optical beams reflected
from the end mirrors $A$ and $B$ and transmitted by the $50\%$ beamsplitter
$C$ on opposite sides of the double highly reflecting (zero transmission)
mirror $D$ (to eliminate linear coupling). In the engineering realization of
this scheme, $A'$ and $B'$ can be rigidly connected to the beamsplitter,
and can be focusing reflectors, making it possible for the mass $m$ of
$D$ to be smaller.

It is easy to see that, due to the beamsplitter, the optical beams from
arms A--C and B--C interfere in the shorter arms such that one of them has
amplitude proportional to $a_1+ia_2$ and the other has amplitude proportional
to $a_2+ia_1$ ($a_{1,2}$ are the complex field amplitudes in the longer arms).
This is valid if the geometrical conditions in Fig.1 are satisfied. As a
result, the ponderomotive force $F_{pond}$ acting on mirror $D$ will be
proportional to:
\begin{equation}
  F_{pond}\propto|a_1+ia_2|^2-|a_2+ia_1|^2 \simeq 4|a_1||a_2|\delta\varphi.
\end{equation}
Provided that the initial optical energy ${\cal E}/2$ in the two arms is
nearly the same (${\cal E}=\hbar\omega_o N=\hbar\omega_o a^2$,
$a=|a_1|=|a_2|$), in a quasistatic approximation, this force will be

\begin{equation}
  F_{pond} \simeq \frac{{\cal E}}{L}\delta\varphi
\end{equation}
Note here that there is no direct linear coupling
between modes in this scheme. In other words, modes in the resonator are
coupled via the $\chi^{(3)}$ nonlinearity resulting from the ponderomotive
effect. Linear coupling is due only to the movement of the mirror $D$. The
shift of $D$ changes the lengths of the shorter arms, changing the
interference conditions on the beamsplitter, which consequently leads to a
redistribution of the optical photons between the two modes.

This scheme realizes indirect QND measurement of the operator

\begin{equation}
  \hat{\cal X}_{\pi/2}=i(\hat a_1^+\hat a_2-\hat a_2^+\hat a_1),
\label{1}
\end{equation}
where $\hat a_{1,2}^+$ and $\hat a_{1,2}$ are the creation and annihilation
operators for two different oscillators with the same frequencies $\omega$.
The operator $\hat{\cal X}_{\pi/2}$ presents a special case of the family of
operators
\begin{equation}
  \hat{\cal X}_{\theta} = \hat a_1^+\hat a_2e^{i\theta}+\hat a_2^ +
    \hat a_1e^{-i\theta},
\end{equation}
which we propose to name crossquadrature operators. These operators commute
with the Hamiltonian of the two modes:

\begin{equation}
  [\hat{\cal X}_{\theta},\hbar\omega(\hat a_1^+\hat a_1+\hat a_2^+\hat a_2)]=0,
  \label{2}
\end{equation}
i.e., they are, indeed, QND variables. The eigenstates of the crossquadrature
operators have the form

\begin{equation}
|N,n\rangle={1\over\sqrt{2^N n!(N-n)!}}(\hat a_1^++\hat a_2^+e^{-i\theta})^n
(\hat a_1+\hat a_2e^{i\theta})^{N-n}|0\rangle,
\label{3}
\end{equation}
where $N$ is the sum of quanta in the system and $n$ is an integer
in the range from 0 to $N$. In this state, each of the $N$ quanta has equal
probability to reside in either arm of the interferometer. However, the
amplitudes of these probabilities for $n$ quanta are orthogonal to those of
the other $N-n$ quanta. Due to this peculiar entanglement between the modes,
we shall call eigenstates of the crossquadrature operator symphotonic quantum
states.

The eigenvalues of the operator $\hat{\cal X}_{\theta}$ are $n-(N-n)=2n-N$,
i.e., measuring the crossquadrature variable, the observer determines the
difference between the two kinds of quanta. Symphotonic states (\ref{3})
are very sensitive to the change of the phase difference in the two oscillators.
As we show in Appendix A, a phase shift leads to a transition between states
with different $n$ (preserving the total number of quanta), that can be
detected by measuring $\hat{\cal X}_{\pi/2}$. The probability of this
transition is equal in the case $\delta\varphi\ll1$ to

\begin{equation}
  p={\delta\varphi^2\over4}(N+2n(N-n)),
\end{equation}
and when $n\simeq N/2$, $p$ tends to unity when
$\delta\varphi \simeq \sqrt{8}/N$,
thus allowing, in principle, the registration of these small phase shifts.

\section{Limitations on the sensitivity}

It is not difficult to show that the finite masses of the mirrors $A'$ and
$B'$, as well as the mass of the beamsplitter $C$, do not influence the
behavior of the system if these masses are substantially greater than the
mass $m$.  We will use a standard linear approximation, in which the optical
field can be represented as the sum of the large classical dimensionless
amplitude ${\cal A}$ and the quantum annihilation operator $\hat a$,
neglecting terms of the order of $\hat a^2$ and higher. We suppose also
that $\tau^*_o$ and relaxation time $\tau^*_m$ of the mass $m$ is very
large in comparision with other characteristic times. In this case, the
equations of motion will have the form:

\begin{eqnarray}
  \frac{d\hat a_1(t)}{dt} &=& \omega_o{\cal A}
    \left(\frac{i\hat x_1(t)-\hat x(t)}{L}+\frac{ih(t)}{2}\right) +
  \displaystyle\int\limits_0^\infty
    \sqrt{\frac{\delta_o}{\pi}}\hat b_1(\omega)e^{-i(\omega+\omega_o)t}
  d\omega  \nonumber \\
  \frac{d\hat a_2(t)}{dt} &=& \omega_o{\cal A}
      \left(\frac{i\hat x_2(t)+\hat x(t)}{L}-\frac{ih(t)}{2}\right) +
  \displaystyle\int\limits_0^\infty
    \sqrt{\frac{\delta_o}{\pi}}\hat b_2(\omega)e^{-i(\omega+\omega_o)t}
  d\omega  \nonumber \\
  m\frac{d^2\hat x(t)}{dt^2} &=& \frac{i\hbar\omega_o{\cal A}}{L}
    \left(\hat a_1^+(t)-\hat a_1(t)+\hat a_2(t)-\hat a_2^+(t)\right) +
    \hat F^{meter}(t) + \hat F^{mech}(t)    \nonumber \\
  M\frac{d^2\hat x_1(t)}{dt^2} &=& \frac{\hbar\omega_o{\cal A}}{L}
    \left(\hat a_1(t)+\hat a_1^+(t)\right)  \nonumber \\
  M\frac{d^2\hat x_2(t)}{dt^2} &=& \frac{\hbar\omega_o{\cal A}}{L}
    \left(\hat a_2(t)+\hat a_2^+(t)\right)
  \label{main_set}
\end{eqnarray}
where $x_{1,2}$ are the displacements of the mirrors $A$ and $B$, $x$ is the
displacement of $D$, $\delta_o=1/2\tau^*_o$ is the decrement of the optical
losses in the resonators, $\hat b_{1,2}(\omega)$ are the corresponding
annihilation operators for the heatbath modes, which satisfy the commutational
relations

\begin{equation}
  [\hat b_{1,2}(\omega),\hat b_{1,2}^+(\omega^\prime)] =
    \delta(\omega-\omega^\prime),
\end{equation}
$F^{meter}$ is the fluctuational reaction of the coordinate meter on the
mirror $D$ with mass $m$, $h(t)/2$ is the relative change of the optical
lengths of the resonators (in the case of a gravitational antenna, this is
the dimensionless metric variation), and $F^{mech}$ is the Nyquist
fluctuational force acting on the mass $m$.

The characteristic equation of this system is:

\begin{equation} p^6+\nu^6=0, \end{equation}
where

\begin{equation}
\nu = \left(\frac{2\omega_o^2{\cal E}^2}{mML^4}\right)^{1/6}.
\label{nu}
\end{equation}
It has roots with positive real parts of the order of $\nu$.  Thus, there
exists in the system a dynamic instability with a characteristic time
$\nu^{-1}$.  To suppress this with a feedback loop, it is necessary to
have

\begin{equation}
  \nu < \omega_{gr}. \label{stability}
\end{equation}

The signal-to-noise ratio is equal to (see Appendix B):

\begin{equation}
  \frac{s}{n} = \frac{\omega_o^2{\cal E}^2}{L^2}
  \displaystyle\int\limits_{-\infty}^\infty
    \frac{\omega^6|h_{\omega}(\omega)|^2}
      { m^2(\nu^6-\omega^6)^2 S_x + 2m\omega^4(\nu^6-\omega^6) S_{xF} +
        \omega^8(S_F+S_m+S_o) }
  \frac{d\omega}{2\pi},
  \label{philim}
\end{equation}
where $h_{\omega}(\omega)$ is the signal spectrum,

\begin{equation}
  S_m=\frac{2\kappa T m}{\tau^*_m}
\end{equation}
is the spectral density of $F^{mech}$ ($\kappa$ is the Boltzmann constant
and $T$ is the temperature),

\begin{equation}
  S_o = \frac{\hbar\omega_o{\cal E}}{L^2\tau^*_o\omega^2}
\end{equation}
is the spectral density of the fluctuational force due to dissipation in the
optical resonators, $S_x$ and $S_F$ are the spectral densities of the additive
noise $x^{meter}$ of the meter and of $F^{meter}$, and $S_{xF}$ is the cross
spectral density of $x^{meter}$ and $F^{meter}$.  The values of $S_F, S_x$ and
$S_{xF}(\omega)$ must obey the Heisenberg inequality \cite{book}:

\begin{equation}
  S_x(\omega)S_F(\omega)-S_{xF}^2(\omega) \ge \frac{\hbar^2}{4} .
\end{equation}

The condition for the detection of a signal can be represented in the form:

\begin{equation}
  h \ge \sqrt{h^2_{meter} + h^2_{mech} + h^2_{opt}},
\end{equation}
where

\begin{equation}
  h_{mech} = \frac{L\omega_{gr}}{{\cal E}\omega_o}
    \sqrt{\frac{2\kappa T m}{\tau_m^*\tau_{gr}}} =
  2\frac{\omega_{gr}}{\nu^3}\sqrt{\frac{2\kappa T}{\tau_m^*\tau_{gr}M}}
  \label{h_m}
\end{equation}
is the limit due to the thermal noise of the mass $m$,

\begin{equation}
  h_{opt}=\sqrt{\frac{1}{\omega_o^2\tau_o^*\tau_{gr}N}}
  \label{h_opt}
\end{equation}
is the limit due to the optical losses ($\tau_{gr}$ is the duration of the
signal), and $h_{meter}$ is the limit due to the quantum noise of the meter.

It is important to note that the limitation (\ref{h_opt}) is also valid
for the previous scheme \cite{optbar}, based on a different principle for
intracavity measurement (this follows from formula (10) of \cite{optbar}).

The value of $h_{meter}$ is determined by the magnitudes of the spectral
densities $S_x(\omega), S_F(\omega)$, and $S_{xF}(\omega)$ and their frequency
dependence. In the case of a plain coordinate meter:

\begin{equation}
  S_x(\omega)=\mbox{const},\quad
  S_F(\omega)=\mbox{const},\quad
  S_{xF}(\omega)=0.
\end{equation}

With regard to limitation (\ref{stability}), values corresponding to optimal
tuning of the meter will be:
\begin{equation}
  S_F = \frac{\hbar m\omega_{gr}^2}{2}, \quad
  S_x = \frac{\hbar}{2m\omega_{gr}^2} .
\end{equation}
The ultimate sensitivity of the meter is determined, in this case, by the
formula
\begin{equation}
  h_{meter} = \frac{L}{\omega_o{\cal E}}
    \sqrt{\frac{\hbar m\omega_{gr}^4}{\tau_{gr}}} =
  \sqrt{2}\left(\frac{\omega_{gr}}{\nu}\right)^3 h_{SQL}(M) .
  \label{h_meter}
\end{equation}

Thus, because of (\ref{stability}), it is impossible in this case
to reach a sensitivity corresponding to $h_{SQL}(M)$.

To preserve a sensitivity at the level of $h_{SQL}(M)$ and lower
the requirements on the energy, one can use an advanced meter providing higher
precision for monitoring the mass $m$. A speed meter \cite{speedmeter} can
be used for this monitoring. This can be realized in the form of an ordinary
parametric electromagnetic displacement transducer (operating at microwave
wavelengths) with an additional buffering cavity, coupled with the main
(working) cavity \cite{speedmeter}. We show in Appendix C that, in this case,

\begin{equation}
  S_x(\omega) =
    \frac{\hbar d^2\Omega_e^4}{4\omega^2\omega_e W_e\sin^2\Phi},
  \quad
  S_F(\omega) =
    \frac{\hbar\omega_e W_e\omega^2}{d^2\Omega_e^4},
  \quad
  S_{xF}(\omega) = -\frac{\hbar}{2}\cot\Phi,
  \label{sp_dens}
\end{equation}
where $\omega_e$ is the microwave frequency, $W_e$ is the microwave pump power,
$d$ is an equivalent parameter with the dimensions of length, which
characterizes the tunability of the transducer \cite{transducer}:

\begin{equation}
  d^{-1} = \frac{1}{\omega_e}\frac{\partial\omega_e}{\partial x} ,
\end{equation}
$\Omega_e$ is the beat frequency between the working and buffering resonators,
which must satisfy the conditions $\Omega_e \gg \omega_{gr}$ and
$\Omega_e/\tau_e^* \gg \omega_{gr}^2$ ($\tau_e^*$ is the relaxation time
due to the coupling with the transmission line), and $\Phi$ is the phase of the
local oscillator used for detection of the microwave signal. For optimal
tuning of the meter parameters, when
\begin{equation}
  W_e =
    \frac{md^2\Omega_e^4}{2\omega_e}\frac{\omega_{gr}^6}{\nu^6}
\end{equation}
and
\begin{equation} \cot\Phi = -\frac{\omega_{gr}^6}{\nu^6}, \end{equation}
the limiting sensitivity will be
\begin{equation} h_{meter} = \sqrt{2} h_{SQL}(M). \end{equation}

If, for example, $\omega_o=2\times 10^{15}s^{-1}$, $L=4\times 10^5cm$,
$\omega_{gr}=10^3s^{-1}$ (these values correspond to the values for the LIGO
antenna \cite{LIGO}), $m=1g$, and ${\cal E}=10^6erg$, then
$\nu\simeq 5\times 10^2s^{-1}$, and condition (\ref{stability}) is satisfied.
If in addition $d=1cm$ (the value achieved in high-Q sapphire disk resonators
\cite{MW_resonator}) and $\Omega_e=3\times 10^3s^{-1}$, then the required
microwave pump power will be $W_e=3\times 10^4 erg/s$. Thus, the analyzed
scheme makes it possible to dramatically decrease the requirements for the
optical circulating energy by using a microwave transducer with a reasonable
set of parameters.

Under these conditions, however, the requirements for the level of dissipation
in the probe mass $m$ increase as the signal that must be registered decreases:

\begin{equation} \Delta v_m \simeq \frac{\nu^3}{\omega_{gr}^3} \Delta v_{SQL}, \end{equation}
where

\begin{equation} \Delta v_{SQL} = \sqrt{\frac{\hbar}{m\tau_{gr}}} \end{equation}
If the above parameters are chosen,
$\Delta v_m \simeq 1/8 \Delta v_{SQL}$.
In order for the dissipation not to deteriorate the sensitivity, it is
necessary that $h_{mech} < h_{meter}$, or

\begin{equation} \frac{2\kappa T}{\tau_m^*} < \frac{\hbar\nu^6}{4\omega_{gr}^4}. \end{equation}
For example, for $T=4K$, $\tau_m^* > 3\times 10^8s$. Thus, the requirements
for the dissipation in the mass $m$ are severe, but achievable
\cite{mitrofanov}.

Losses in the optical resonator will not influence the sensitivity
if $h_{opt} < h_{meter}$, which is equivalent to the condition
\begin{equation} \tau_o^* > \frac{{\cal E}_{SQL}}{{\cal E}\omega_{gr}} , \end{equation}
or, for the parameter values introduced above, $\tau_o^* > 1s$.
This is quite possible in a LIGO-type detector with optical mirrors
available today.

\section{Comparison with the ``optical bar'' scheme}

In \cite{optbar}, not all regimes for the ``optical bar'' scheme were
analyzed in detail. Moreover, there was unfortunately an error in the formula
following formula (12) (term ``1'' under the root should be omitted).

Here, we shall limit our treatment to ``wideband'' regimes, when the range of
the signal frequencies is far from the resonant frequencies in the
signal-to-noise integral; these regimes are the most useful from the practical point
of view. The ``narrowband'' regime, in which it is possible to attain
sensitivity better than the SQL, has already been considered in detail in
\cite{optbar}. In our analysis, we shall assume that
$\omega_{gr} < \Omega$ ($\Omega$ is the beat frequency in the system of
two coupled optical resonators; the case of $\omega_{gr} > \Omega$ is
difficult to realize in practice, and does not provide any interesting new
results) and $m\ll M$. The behavior of the system is determined by the
parameter with the dimensions of frequency
\begin{equation}
\Theta=\left(\frac{2\omega_o{\cal E}\Omega}{L^2}
\left(\frac{1}{m}+\frac{1}{2M}\right)\right)^{1/4}
\end{equation}
This frequency describes the influence of the ponderomotive force on the
dynamics of the system, and plays a role analogous to $\nu$ (see formula
(\ref{nu})).

It is possible to distinguish three cases, depending on the level of the
circulating energy (the value of $\Theta$).

\subsection{Weak pump power, $\Theta^2 < \omega_{gr}\Omega$}

If a plain coordinate meter is used, the calculations give the following
result:

\begin{equation}
  h_{meter} = \left(\frac{\omega_{gr}\Omega}{\Theta^2}\right)^2h_{SQL}(m)
    > h_{SQL}(m),
\end{equation}
where
\begin{equation}
  h_{SQL}(m)=\frac{1}{L}\sqrt{\frac{\hbar}{m\omega^2_{gr}\tau_{gr}}}
  \label{sql_m}
\end{equation}
If a speed meter is used, the sensitivity can be higher:

\begin{equation}
  h_{meter} = \frac{\omega_{gr}\Omega}{\Theta^2} h_{SQL}(m)
    > h_{SQL}(m),
\end{equation}
but is still lower than even $h_{SQL}(m)$.

\subsection{Intermediate case, $\omega_{gr}\Omega < \Theta^2 <
  \omega_{gr}\Omega\sqrt{2M/m}$}
For a plain coordinate sensor, the best sensitivity in this case is

\begin{equation} h_{meter} = \frac{\omega_{gr}\Omega}{\Theta^2}h_{SQL}(m), \end{equation}
i.e., $h_{meter}$ is smaller than $h_{SQL}(m)$, and $h_{meter} \to
h_{SQL}(M)$ if $\Theta^2\to\omega_{gr}\Omega\sqrt{2M/m}$.
The required optical energy in this case is

\begin{equation}
{\cal E}=\frac{\Omega}{\omega_{gr}}{\cal E}_{SQL} > {\cal E}_{SQL}.
\end{equation}
The use of a speed meter in this regime does not give a gain in sensitivity,
however an increase in sensitivity is possible if an advanced coordinate
detector with correlated noises is used ($S_{xF} \ne 0$) \cite{Syrtsev}. In
this case

\begin{equation} h_{meter} = \left(\frac{\omega_{gr}\Omega}{\Theta^2}\right)^2h_{SQL}(m), \end{equation}
if $\Theta^2 < \omega_{gr}\Omega\sqrt[4]{2M/m}$, and
\begin{equation} h_{meter} = h_{SQL}(M)\end{equation}
otherwise. The required energy in the latter case can be lower than
${\cal E}_{SQL}$, but with respect to a possible dynamical instability, which
appears when $\Theta^2 \ge \Omega^2/4$:

\begin{equation} {\cal E}>\sqrt[8]{\frac{8m}{M}}{\cal E}_{SQL} \end{equation}

\subsection{Strong pump power, $\Theta^2 \gg \omega_{gr}\Omega\sqrt{2M/m}$}

This is the ``optical bar'' regime, when the masses $M$ and $m$ move together,
and are connected by electromagnetic rigidity. In this case, a plain coordinate
meter provides a sensitivity corresponding to the standard quantum limit
$h_{SQL}(M)$, and use of a speed meter makes it possible to overcome this
limit, but with higher energy:

\begin{equation}h_{meter} = \frac{\omega_{gr}\Omega}{\Theta^2}h_{SQL}(m)=
\sqrt{ \frac{{\cal E}_{SQL}\Omega}{{\cal E}\omega_{gr}}}  h_{SQL}(M)\end{equation}

Note that, in this case, also, the total mass $\sim 2M$ is present in the
expression for the thermal limit. This result is quite understandable,
since, in this regime, thermal fluctuations of the small mass $m$ act on the
large compound mass $2M+m$.

\section{Conclusion}
Quantum mechanics sets severe limits on the sensitivity and the required
circulating energy in traditional free-mass gravitational-wave antennas.
One possible way to beat these limits is to use intracavity QND measurements.
In this paper, we have analyzed a new QND observable and its corresponding
symphotonic quantum states, which possess a number of features that make
it promising for experiments requiring registration of small phase variations:

1) Unlike other known QND observables, this one is a joint integral of
motion for two quantum oscillators with equal frequencies.

2) The crossquadrature observable is very sensitive to the phase difference of
the oscillators.
Phase differences of the order of $1/N$ (the theoretical limit for phase
measurements) can be detected, where $N$ is number of quanta in the system.

3) Well-known methods for the QND measurement of electromagnetic energy can
be used to measure this new observable.

We have considered a practical optical scheme in which the new observable
can be used for the detection of gravitational waves. Our estimates show
that, in combination with advanced coordinate meters, this scheme provides
a sensitivity of the same order as that for planned antennas at significantly
lower energies.

Summarizing the results of this article and of
\cite{NL96,optbar,Levin}, we conclude that intracavity measurements with
automatically organizing nonclassical optical quantum states make it possible,
in principle, to lower the required power levels and in several cases to
achieve sensitivity better than the standard quantum limit.

We note also that the schemes we have analyzed do not cover all possible
geometries for intracavity measurements with ponderomotive nonlinearity.
Better realizations with higher responses are probably possible.

\acknowledgments
We thank Prof. Kip S.Thorne and Yuri Levin for stimulating discussions.
This research was partially supported by the California Institute of
Technology, US National Science Foundation and the Russian Foundation for
Basic Research (grant \#96-15-96780).

\newpage

\appendix
\section{The evolution of a symphotonic state}
The evolution operator describing the phase shifts $\phi_1$ in the first mode
and $\phi_2$ in the second one is equal to

\begin{equation}
  \hat U(\phi_1,\phi_2) = \exp\left(\frac{\phi_1\hat n_1+\phi_2\hat n_2}
    {i\hbar}\right)
\end{equation}
where $\hat n_{1,2}$ are the operators for the number of quanta in the modes.
Hence

\begin{equation}
  \hat U(\phi_1,\phi_2) \hat a_1^+ \hat U^+(\phi_1,\phi_2) =
    \hat a_1^+ e^{-i\phi_1},
\end{equation}
\begin{equation}
  \hat U(\phi_1,\phi_2) \hat a_2^+ \hat U^+(\phi_1,\phi_2) =
    \hat a_2^+ e^{-i\phi_2},
\end{equation}
and

\begin{equation} \hat U(\phi_1,\phi_2)|0\rangle = |0\rangle . \end{equation}
Taking into account formula (\ref{3}) and omitting the unimportant factor
$e^{-\frac{i(\phi_1+\phi_2)N}{2}}$, we can obtain:

\begin{equation}
  \hat U(\phi_1,\phi_2)|N,n\rangle =
    \frac{1}{\sqrt{2^N n!(N-n)!}}
    (\hat A^+\cos\delta\phi + iB^+\sin\delta\phi)^n
    (\hat B^+\cos\delta\phi + iA^+\sin\delta\phi)^{N-n}|0\rangle
\end{equation}
where
\begin{equation}
  \hat A^+ = a_1^+ + a_2^+e^{-i\theta}, \qquad
  \hat B^+ = a_1^+ - a_2^+e^{-i\theta}
\end{equation}
and

\begin{equation} \delta\phi = \phi_2 - \phi_1. \end{equation}
If $\delta\phi \ll 1$ then

\begin{eqnarray}
  & & \hat U(\phi_1,\phi_2)|N,n\rangle \simeq
  \left(1-\frac{\delta\phi^2}{8}(N+2n(N-n))\right)|N,n\rangle +  \nonumber\\
  & & i\delta\phi\left(
    \sqrt{n(N-n+1)}|N,n-1\rangle + \sqrt{(n+1)(N-n)}|N,n+1\rangle
  \right) -                                                     \nonumber\\
  & &
  \frac{\delta\phi^2}{2}\left(\sqrt{n(n-1)(N-n+1)(N-n+2)}|N,n-2\rangle +
                                                  \right.     \nonumber\\
  & & \left. \sqrt{(n+1)(n+2)(N-n)(N-n-1)}|N,n+2\rangle \right)
\end{eqnarray}
Thus, the probability for changing the number $n$ is equal to

\begin{equation}
  p = 1-|\langle N,n|\hat U(\phi_1,\phi_2)|N,n\rangle|^2 \simeq
    \frac{\delta\phi^2}{4}(N+2n(N-n)) .
  \label{A}
\end{equation}

\section{Signal-to-Noise ratio}
Equations (\ref{main_set}) can be rewritten in the form:

\begin{eqnarray}
    \frac{d\hat{\cal N}(t)}{dt} &=& -\frac{2\omega_o N}{L}\hat x(t) +
      {\cal A}\displaystyle\int\limits_0^{\infty}
      \sqrt{\frac{\delta_o}{\pi}}\left(
        (\hat b_1^+(\omega)-\hat b_2^+(\omega))e^{i(\omega+\omega_o)t}
      + h.c.\right)d\omega  \nonumber\\
    \frac{d\hat{\cal X}_{\pi/2}(t)}{dt} &=&
      \frac{2\omega_o N}{L}\hat y(t) + \omega_o Nh(t) +
      {\cal A}\displaystyle\int\limits_0^{\infty}
      i\sqrt{\frac{\delta_o}{\pi}}\left(
        (\hat b_1^+(\omega)-\hat b_2^+(\omega))e^{i(\omega+\omega_o)t}
      + h.c.\right)d\omega  \nonumber\\
    m\frac{d^2 x(t)}{dt^2} &=&
      \frac{\hbar\omega_o}{L}\hat{\cal X}_{\pi/2}(t)+
      \hat F^{meter}(t) \nonumber\\
    2M\frac{d^2\hat y(t)}{dt^2} &=&
      \frac{\hbar\omega_o}{L}\hat{\cal N}(t)
  \label{B}
\end{eqnarray}
where
\begin{equation}
  \hat{\cal N}={\cal A}(\hat a_1+\hat a_1^+-\hat a_2-\hat a_2^+),\qquad
  \hat{\cal X}_{\pi/2}=i{\cal A}(\hat a_2-\hat a_2^+-\hat a_1+\hat a_1^+),
\end{equation}
$\hat{\cal N}$ - difference of number of quanta in the two arms,
$y = (x_1-x_2)/2$, and $h.c.$ stands for Hermitian conjugation.
Hence, the spectrum of $x(t)$ is equal to

\begin{equation} x(\omega) = x_{signal}(\omega) + X(\omega), \end{equation}
where

\begin{equation}
  x_{signal}(\omega) = \frac{\hbar\omega_o^2N}{mL}
    \frac{-i\omega^3}{\nu^6-\omega^6}h(\omega)
\end{equation}
is the signal spectrum,

\begin{equation}
  X(\omega) = \frac{\omega^4}{m(\nu^6-\omega^6)}
    \left(F^{meter}(\omega)+F^{mech}(\omega)+F^{opt}(\omega)\right)
\end{equation}
is the spectrum of fluctuations of $x(t)$, and $F^{opt}(\omega)$ is the
spectrum of force

\begin{equation}
  F^{opt}(t) = \frac{\hbar\omega_o{\cal A}}{L}
    \displaystyle\int\limits_0^{\infty}\sqrt{\frac{\delta_{opt}}{\pi}}\left[
      \left(\frac{1}{i\omega}+\frac{\omega_o{\cal E}}{ML^2\omega^4}\right)
      \left(\hat b_1^+(\omega)-\hat b_2^+(\omega)\right)
        e^{i(\omega+\omega_o)t}d\omega + h.c.
    \right]
\end{equation}
The output signal of the coordinate meter is equal to

\begin{equation}  \tilde x(t) = x(t) + x^{meter}(t) , \end{equation}
where $x^{meter}(t)$ is the additive noise of the meter. Hence,

\begin{equation}
  \tilde x(\omega) = x(\omega) + x^{meter}(\omega)
    = x_{signal}(\omega) + X(\omega) + x^{meter}(\omega) ,
\end{equation}
and the SNR is equal to (\ref{philim}).

\section{Microwave speed meter}

Let us consider two coupled microwave resonators. The first is connected
to an output waveguide (see Fig.2), and it's eigenfrequency depends
on the coordinate $x$ to be measured:

\begin{equation} \tilde\omega(x) = \omega_e\left(1-\frac{x}{d}\right), \end{equation}
while the second is pumped by the resonant power $U_0\cos\omega_et$.

The equations of motion for such a system are

\begin{eqnarray}
  & & \frac{d^2 q_1(t)}{dt^2} + 2\delta_e\frac{d q_1(t)}{dt^2} +
    \omega_e^2\left(1-\frac{x(t)}{d}\right)^2 q_1(t) =
  2\omega_e\Omega_e q_2(t) + \frac{2\omega_e}{\rho}U^{fluct}(t) \nonumber \\
  & & \frac{d^2 q_2(t)}{dt^2} + \omega_e^2 q_2(t) =
  2\omega_e\Omega_e q_1(t) + \frac{\omega_e}{\rho}U_0\cos\omega_e t ,
\end{eqnarray}
where $q_{1,2}$ are the generalized coordinates of the resonators, $\rho$ is
the wave impedance of the resonators, $\delta_e = 1/2\tau_e^*$, $\tau_e^*$
is the relaxation time of loaded first resonator, and $U^{fluct}$ are
fluctuations in the waveguide (we neglect intrinsic losses and corresponding
fluctuations of the resonators).

Linearizing these equations in the strong-pumping approximation and using
the method of slowly varying amplitudes, we can obtain:

\begin{eqnarray}
  \frac{d a_1(t)}{dt} + \delta_e a_1(t) & = &
    - \Omega_e b_2(t) - \frac{U_s^{fluct}(t)}{\rho} \nonumber \\
  \frac{d b_1(t)}{dt} + \delta_e b_1(t) & = &
    \frac{\omega_e q_0}{d}x(t) +
    \Omega_e a_2(t) + \frac{U_c^{fluct}(t)}{\rho}   \nonumber \\
  \frac{d a_2(t)}{dt} & = & - \Omega_e b_1(t)       \nonumber \\
  \frac{d b_2(t)}{dt} & = & \Omega_e a_1(t)
\end{eqnarray}
where $a_{1,2}$ and $b_{1,2}$ are the amplitudes of the cosine and sine
quadrature components of $q_{1,2}$, $U_{c,s}^{fluct}$ are the same for
$U^{fluct}$, and $q_0$ is the mean value of the amplitude of oscillations in
the first resonator.

Solution of these equations in the spectral representation gives:

\begin{equation}
  a_1(\omega) = -\frac{i\omega U_s^{fluct}(\omega)}{\rho{\cal L}(\omega)},
  \qquad
  b_1(\omega) = \frac{i\omega}{{\cal L}(\omega)}\left(
    \frac{\omega_e q_0}{d}x(\omega) + \frac{U_c^{fluct}(\omega)}{\rho}\right),
  \label{C1}
\end{equation}
where ${\cal L}(\omega) = \Omega_e^2-\omega^2+i\omega\delta_e$.
The output wave in the waveguide can be represented in the form:

\begin{eqnarray}
  & & U^{out}(t) = U^{fluct}(t) - \frac{2\delta_e\rho}{\omega_e}\frac{d q_1(t)}{dt} =
                                                               \nonumber \\
  & & \left(U_c^{fluct}(t) - 2\delta_e\rho b_1(t)\right)\cos\omega_e t +
  \left(U_s^{fluct}(t) + 2\delta_e\rho a_1(t)\right)\sin\omega_e t
\end{eqnarray}
If a homodine detector with $U_{LO}\propto sin(\omega_e t+\Phi)$ is used,
where $\Phi$ is the phase of the local oscillator, the output signal of the
detector is proportional to

\begin{equation}
  \tilde U(t) = \left(U_c^{fluct}(t) - 2\delta_e\rho b_1(t)\right)\sin\Phi +
    \left(U_s^{fluct}(t) + 2\delta_e\rho a_1(t)\right)\cos\Phi
\end{equation}
Substitution into this expression of the solution (\ref{C1}) gives
that the spectrum of $\tilde U$ is equal to
\begin{equation}
  \tilde U(\omega) =
    -\frac{2i\omega\omega_e\delta_e\rho q_0 \sin{\Phi}}{{\cal L}(\omega)d}
    \left(x(\omega)+x^{meter}(\omega)\right),
\end{equation}
where
\begin{equation}
  x^{meter}(\omega) = -\frac{d}{2i\omega\omega_e\delta_e q_0\rho\sin\Phi}
    \left(i\omega(\Omega_e^2-\omega^2-i\omega\delta_e)\right)
    \left(U_c^{fluct}(\omega)\sin\Phi + U_s^{fluct}(\omega)\cos\Phi\right)
  \label{x_fl}
\end{equation}
is the spectrum of the additive noise of the meter.

The fluctuational reaction force of the meter is equal to
\begin{equation}
F^{meter}(t) = \frac{q_0\rho\omega_e}{d}a_1(t),
\end{equation}
or, in spectral form,

\begin{equation}
  F^{meter}(\omega) =
    -\frac{i\omega\omega_e q_0 U_s^{fluct}(\omega)}{{\cal L}(\omega)d}
  \label{F_fl}
\end{equation}
If the frequency $\omega$ is relatively small:

\begin{equation}
  \omega^2 \ll \Omega_e^2 \mbox{\quad and \quad}
    \omega\delta_e \ll \Omega_e^2
\end{equation}
then expressions (\ref{F_fl},\ref{x_fl}) directly give the spectral densities
(\ref{sp_dens}).

\newpage

\begin{figure}

\unitlength=2pt

\begin{picture}(180,140)

\thinlines                                   %beams
\put(14,55){\line(1,0){116}}
\put(55,14){\line(0,1){116}}
\put(14,55){\line(1,-1){41}}

\thicklines                                  %mirror A
\put(45,130){\line(1,0){20}}
\put(70,130){\makebox(0,0)[cc]{\large{A}}}
\put(57,100){\makebox(0,0)[lc]{$L-l$}}

\thicklines                                  %mirror A'
\put(45,10){\line(5,2){20}}
\put(70,15){\makebox(0,0)[cc]{\large{A'}}}
\put(57,32){\makebox(0,0)[lc]{$l+\lambda/4$}}

\thicklines                                  %mirror B
\put(130,45){\line(0,1){20}}
\put(130,70){\makebox(0,0)[cc]{\large{B}}}
\put(100,57){\makebox(0,0)[cb]{$L-l-\lambda/4$}}

\thicklines                                  %mirror B'
\put(10,45){\line(2,5){8}}
\put(15,70){\makebox(0,0)[cc]{\large{B'}}}
\put(32,57){\makebox(0,0)[cb]{$l$}}

\thicklines                                  %mirror C
\put(45,45){\line(1,0){20}}\put(65,45){\line(0,1){20}}
\put(45,45){\line(0,1){20}}\put(45,65){\line(1,0){20}}
\put(45,45){\line(1,1){20}}
\put(70,70){\makebox(0,0)[cc]{\large{C}}}

\thicklines                                  %mirror D
\put(29,29){\line(1,1){11}}
\put(25,25){\makebox(0,0)[cc]{\large{D}}}
\end{picture}
\caption{The scheme of measurement of the crossquadrature observable}
\end{figure}
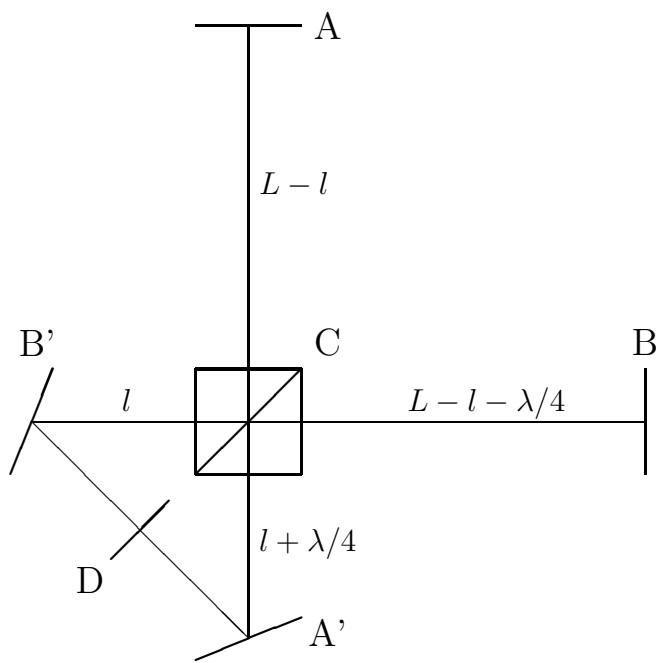

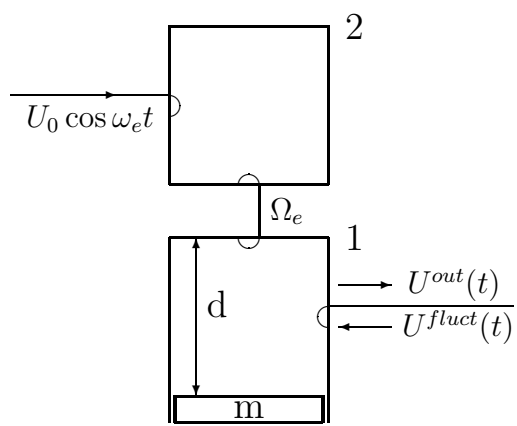
\begin{figure}
\unitlength=2pt

\begin{picture}(180,120)

\thicklines      % resonator 2
\put(45,55){\line(1,0){30}}\put(75,55){\line(0,1){30}}
\put(45,55){\line(0,1){30}}\put(45,85){\line(1,0){30}}
\put(80,85){\makebox(0,0)[cc]{\large{2}}}

\thinlines       % input
\put(15,72){\line(1,0){30}}\put(45,70){\oval(4,4)[r]}
\put(25,72){\vector(1,0){10}}
\put(30,70){\makebox(0,0)[ct]{$U_0\cos\omega_e t$}}

\thicklines      % resonator 1
\put(75,10){\line(0,1){35}}
\put(45,10){\line(0,1){35}}\put(45,45){\line(1,0){30}}
\put(80,45){\makebox(0,0)[cc]{\large{1}}}

\thicklines      % m
\put(46,10){\line(1,0){28}}\put(74,10){\line(0,1){5}}
\put(46,10){\line(0,1){5}}\put(46,15){\line(1,0){28}}
\put(46,10){\makebox(28,5)[cc]{\large{m}}}

\thinlines
\put(50,30){\vector(0,-1){15}}\put(50,30){\vector(0,1){15}}
\put(52,30){\makebox(28,5)[lc]{\large{d}}}

\thinlines       % output
\put(75,32){\line(1,0){35}}\put(75,30){\oval(4,4)[l]}
\put(77,36){\vector(1,0){10}}
\put(90,36){\makebox(0,0)[lc]{$U^{out}(t)$}}
\put(87,28){\vector(-1,0){10}}
\put(89,28){\makebox(0,0)[lc]{$U^{fluct}(t)$}}

\thinlines       % coupling
\put(62,45){\line(0,1){10}}
\put(60,45){\oval(4,4)[b]}
\put(60,55){\oval(4,4)[t]}
\put(64,50){\makebox(0,0)[lc]{$\Omega_e$}}

\end{picture}
\caption
{The scheme of microwave speedmeter}
\end{figure}

\newpage

\centerline{Figure captions}
\bigskip
\bigskip

Figure 1. The scheme of measurement of the crossquadrature observable

\bigskip
\bigskip

Figure 2. The scheme of microwave speedmeter

\bigskip
\bigskip

\end{document}